\documentclass[twocolumn,showpacs,preprintnumbers,amsmath,amssymb]{revtex4}
\usepackage{psfrag}
\usepackage{graphicx}
\usepackage{bm}
\usepackage{setspace}

\usepackage{xcolor}
\usepackage{epsfig}

\newcommand{\be}{\begin{equation}}
\newcommand{\ee}{\end{equation}}

\newcommand{\xx}{\mathbf{x}}

\newcommand{\UU}{\mathbf{U}}

\newcommand{\vor}{\mbox{\boldmath{$\omega$}}}

\begin{document}

\preprint{Under revision for submission to PHYSICA D}
\title{Survival distribution of the stretching and tilting of vortical structures in isotropic turbulence. Anisotropic filtering analysis.
}

\author{Daniela Tordella$\natural$}
\email{daniela.tordella@polito.it}
\author{Luca Sitzia$\sharp$}
\author{Silvio Di Savino$\natural$}
\affiliation{$\natural$ Dipartimento di Ingegneria Meccanica e Aerospaziale, Politecnico di Torino, Corso Duca degli Abruzzi 24, 10129
Torino, Italy\\$\sharp$ Dottorato in Statistica e Matematica Applicata, Scuola di Dottorato "V.Pareto" - Universita' di Torino, via Real Collegio 30, 10024 Moncalieri (To), Italy
}
%



\date{\today}

\begin{abstract}

-----------------------------------------------------------------------------------------------------------------------------
\\
Using a Navier-Stokes isotropic turbulent field numerically simulated in a box with a discretization of $1024^3$ (Biferale L. et al. {\it Physics of Fluids}, {\bf 17}(2), 021701/1-4 (2005)), we show that
the probability of having a stretching-tilting larger than twice the local enstrophy is negligible.
By using an anisotropic kind of filter in the Fourier space, where  wavenumbers that have at least one component below a threshold or inside a range are removed, we analyze these survival statistics when the large, the small inertial or the small inertial and dissipation scales are filtered out.
It can be observed that, in the unfiltered isotropic field, 
the probability of the ratio  ($| \vor\cdot\nabla\UU |/|\vor|^2 $) being higher than a given  threshold  is higher  than in the fields where the large scales were filtered out. At the same time,  it is lower than  in the fields were the small inertial and dissipation range of scales is filtered out. This is basically due to the suppression of compact structures in the ranges that have been filtered in different ways. The partial removal of the background of filaments and sheets does not have a first order effect on these statistics.
These results are discussed in the light of a hypothesized relation between vortical filaments, sheets and blobs in physical space and in Fourier space. The study in fact can be viewed as a kind of test for this idea and tries to highlight its limits.
We conclude that a qualitative relation in physical space and in Fourier space can be supposed to exist for blobs only. That is for the near isotropic structures which are sufficiently described by  a single spatial scale and  do not suffer from the disambiguation problem as filaments and sheets do.


Information is also given on the filtering effect on statistics concerning the inclination  of the strain rate tensor eigenvectors with respect to vorticity. In all filtered ranges, eigenvector 2 reduces its alignment, while eigenvector 3 reduces its misalignment.  All filters increase the gap between the most extensional eigenvalue $<\lambda_1>$ and the intermediate one $<\lambda_2>$ and the gap between this last $<\lambda_2>$ and the contractile eigenvalue $<\lambda_3>$. When the large scales are missing, eigenvalue modulus 1 and 3 become nearly equal,  similar to the modulus of the related components of the enstrophy production.

-----------------------------------------------------------------------------------------------------------------------------

\end{abstract}

\maketitle

\section{\label{introduction}Introduction}

The  formation of  spatial and temporal internal scales can in
part be   associated to the
stretching and tilting of vortical structures.
Many aspects of the behavior of turbulent fields have been
associated to this phenomenon: the onset of instability,
vorticity intensification or damping, or the
three-dimensionalization of the flow field 
\cite{my71-75, tl72, Pope}.
In the standard picture of turbulence,
the energy cascade to smaller scales is interpreted in terms of the
stretching of vortices due to the interaction with similar eddy
size (see for example  \cite{Frisch}). A number of statistical details on the stretching phenomenon and the closely related enstrophy production can be found in the monography by Tsinober (2001, see in particular Chapter 6, \cite{Tsinober_book1}).

Although  the important physical role of these inertial phenomena  is
recognized, the literature does not often
include statistical information on quantities such as
the magnitude or the components of $\vor\cdot\nabla\UU $.
For instance, in a letter to Nature (2003) dedicated to the measurements of  intense rotation and
dissipation in turbulent flows, Zeff et al. \cite{Zeff}  observe that the understanding of the temporal interactions
between stretching and vorticity is crucial to the science of extreme
events in turbulence. However, the statistics  presented there concern dissipation and enstrophy and not directly  stretching.
The literature more often includes statistical information concerning other gradient quantities such as
the strain rate or the rate-of-rotation tensors, and, in particular, their fundamental constituents: the longitudinal or transverse velocity derivatives.
Over the last 20 years, statistics on the skewness and flatness factors of the velocity
derivative have been considered  in a
number of laboratory and numerical studies that show how these
quantities increase monotonically with the Reynolds number, see e.g.
\cite{Tsinober_92_jfm} and the review by Sreenivasan  and Antonia
(1997)\cite{sa97}.

In the case of turbulent wall flows, laboratory measurements of both
the mean and the r.m.s. of fluctuations of the stretching components
across the two-dimensional boundary layer have been reported by
Andreapoulos and Honkan (2001) \cite{ah01}. In this study, the
normalized r.m.s values of the stretching components are very
significant throughout the boundary layer  and reach values that are one order
of magnitude larger than the mean span-wise component (the only
significant mean component, however  and only in the near wall
region). The  values observed for the r.m.s. of the stretching range
from 0.04, close to the wall, to about 0.004 in the outer part.

In a study concerning the structure and dynamics of vorticity and rate
of strain in incompressible homogeneous turbulence, Nomura and Post (1998),\cite{Nomura}
demonstrate the significance of both local  dynamics
(influence of local vorticity) and spatial structure (influence through non-local pressure Hessian) in the
interaction of the vorticity and strain rate tensor. The behaviour of high-amplitude
rotation-dominated events cannot be solely represented by local dynamics due to the
formation of distinct spatial structure. Instead, high-amplitude strain dominated
regions are generated predominantly by local dynamics. The associated
structure is less organized and more discontinous than the one associated with rotation dominated
events. They conclude that non-local effects are significant in the dynamics of small scale
motion. This should be considered in the interpretation of single-point statistics.
Characterizations of small-scale turbulence should consider not only the typical structures
there present but also typical structure interactions. In this context these authors offer the radial distribution of the magnitude of the strain rate tensor normalized on the enstrophy. In this paper the maximum value of this magnitude is found  close to 0.2.

Laboratory statistical information on the stretching of field lines
can be found in \cite{ag05}. Here, probability density functions  of the logarithm of the local stretching in N cycles were obtained for several two-dimensional time-periodic confined flows exhibiting chaotic advection.  The stretching fields were observed to be highly correlated in space when N is large, and the probability  distributions were observed to be similar for different flows.

However, a few examples in literature can also be cited regarding direct results for stretching-tilting
statistics. For instance, recently experimental and numerical confirmation has been found of the  predominance of three dimensional turbulent vortex stretching in the positive net enstrophy production. These aspects have been extensively considered in Tsinober (2000) \cite{Tsinober} and in the 2001 monography \cite{Tsinober_book1}, where a number of statistical geometrical details concerning the vortex alignement, compression, tilting, and folding  are outlined.
Through two papers, Constantin, Procaccia and Segel (1995) \cite{Constantin},  Galanti, Procaccia and Segel  (1996) \cite{Galanti} consider the stretching and its relationships with the amplification of vorticity and the straightening of the vortex lines. They show that the same stretching that amplifies the vorticity also tends to straighten out the vortex lines. They also show that in well-aligned vortex tubes, the self-stretching rate of the vorticity is proportional to the ratio of the vorticity and the radius of curvature. In this context,\cite{Galanti} gives statistics on the stretching and vortex line curvature. Numerically this is seen as the appearance of high correlations between the stretching and the straightness of the vortex lines. Regarding to this issue, an important universal feature of fully developed turbulent flows is the preferential alignment of vorticity along the eigen direction of the intermediate eigenvalue of the strain-rate tensor. A number of works both experimental and numerical studies on this result are available (Tsinober, Kit and Dracos JFM (1992), \cite{Tsinober_92_jfm}, Kholmyansky, Tsinober and S. Yorish PoF (2001), \cite{Kolmyansky},  Gulitski et al. JFM (2007 a,b,c), \cite{Gulitski_a,Gulitski_b,Gulitski_c} and Chevillard et al. (2008), \cite{Chevillard}). It should be noticed, however, that in the case of {\it nonlocal} strain rate, Hamlington, Schumacher and Dahm \cite{hsd08}, have observed  a direct assessment of vorticity alignment with the most extensional eigenvector by using data from highly resolved direct numerical simulations.

In the present study, for the case of isotropic turbulence ($Re_{\lambda}=280$, \cite{bbclt05}), we consider statistics related to the intensity of the stretching term in the equation for vorticity.
If we consider the general instantaneous local intrinsic anisotropy of turbulent fields,  looking at stretched structures as filaments and sheets, we would like to be able to disentangle them to follow and understand better their evolution and detailed dynamics. Isotropic filtering  is unable to carry out this job.

 We have conceived a probe function, the ratio between the magnitude of the vortex stretching and the enstrophy, to empirically and statistically  measure the local activity of the stretching phenomenon (see section II).  In addition, we propose an alternative to the commonly used isotropic filter: the cross filter. This is a new empirical, and at the moment limited, attempt to introduce an anisotropic filtering. In section III, we analyze the survival function of the normalized stretching by using the cross filter acting directly on the velocity Fourier space. We do this in the hope of qualitatively  highlighting aspects related to the role of the three-dimensional structures  known as blobs, sheets and filaments and their hypothetical Fourier counterparts. 
 This study can be viewed as a kind of test for this  idea and tries to highlight its limits. Concluding remarks are made in section IV.

\section{\label{function}The normalized stretching-tilting  function}

With reference to the phenomena described by the inertial
nonlinear nonconvective part of the vorticity transport equation,  let us
introduce  a local measure of the process of three-dimensional inner
scales formation

\begin{equation}
f(\xx, t)=\frac{| \vor\cdot\nabla\UU |}{|\vor|^2}(\xx, t)= \frac{| \vor\cdot S_{i,j}|}{|\vor|^2}(\xx, t) .
\end{equation}

\noindent where  $\UU$ 
is the velocity  field, $S_{i,j}$ is the strain rate tensor,  and $\vor$ 
is the vorticity vector. 
The numerator,  the so called
stretching-tilting term of the vorticity equation, is
zero in two-dimensional flows. In 3 D fields, it is commonly believed to be responsible for
the transfer of the kinetic energy from larger to smaller scales (positive or extensional stretching) and viceversa (negative or compressional stretching). According to  definition (1), $f$ depends on the local instantaneous  velocity  and vorticity fields. 
In this study, we leave aside the peculiarity associated to the convective forcing and focus on the action of the fluctuation field only. For simplicity, we consider here the fluctuation of an homogeneous isotropic turbulent field (\cite{bbclt05}). 
Since the stretching term  plays an important role in the enstrophy production, in the previous  definition the normalization by $|\vor|^2$ was adopted. It should be recalled that the square of the vorticity magnitude is the only invariant of the rate of rotation tensor which is non zero and is also the square of the Frobenius norm, an invariant norm of the rate of strain tensor. For this reason, we considered the enstrophy a good candidate as reference quantity  for the product $\vor\cdot\nabla\UU$. In fact, as it can be seen below, the survival probability distribution function of $f$ is very small for values larger than $O(1)$.

Function $f$ was evaluated over a fully resolved
homogeneous isotropic incompressible  steady in the mean turbulence in order to look for  the typical range of values of $f(\xx)$ and to relate
them to the behavior of the various turbulence scales present in an
isotropic field.

\begin{figure}
\psfig{figure=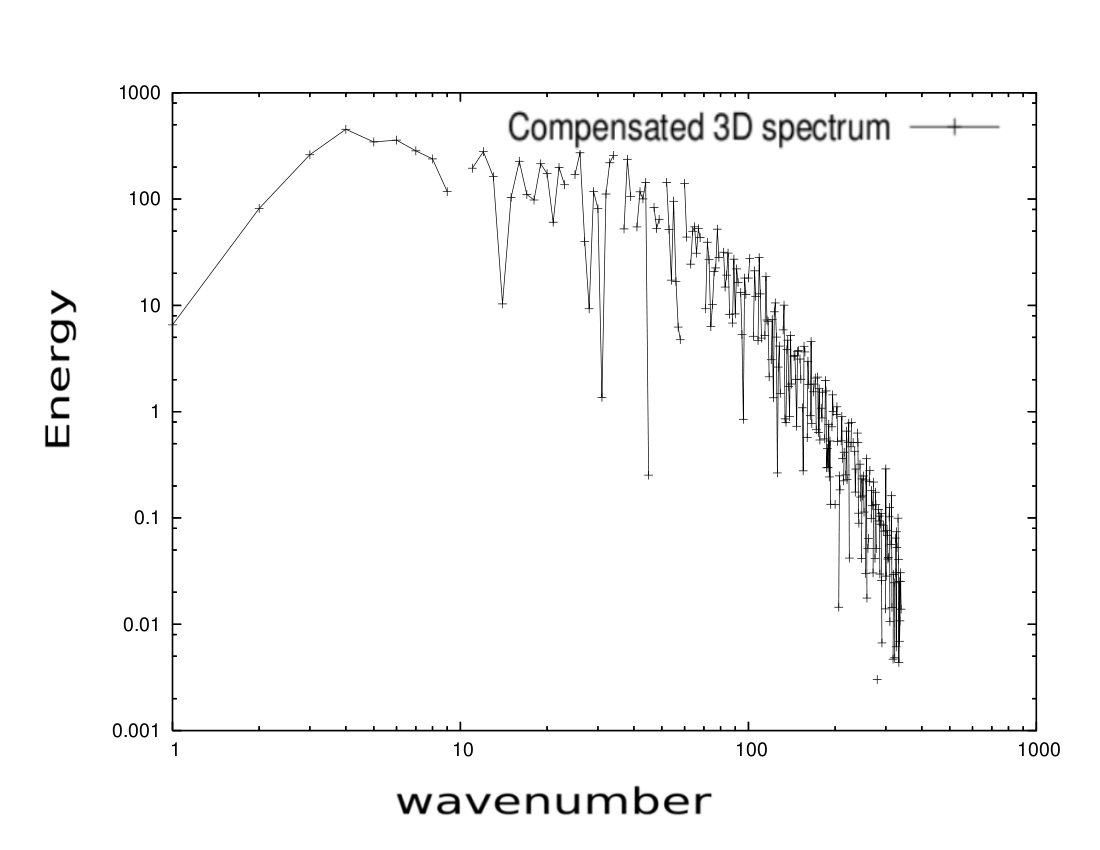,width=1\columnwidth}
\caption{Compensated 3D energy spectrum of one time instant of the turbulent isotropic
field here considered. Open access database http://mp0806.cineca.it/icfd.php. Navier-Stokes
direct numerical simulation in a box with a discretization of $10243$, $Re_{\lambda} = 280$. See e.g.
Biferale L. et al. Physics of Fluids, 17(2), 021701/1-4 (2005).
 }
\label{spe3_compensated}
\end{figure}

The dataset consists of $1024^3$ resolution grid point Direct
Numerical Simulation (DNS) of an isotropic Navier-Stokes forced
field at Reynolds $Re_{\lambda}=280$ \cite{bbclt05}. Each instant in the simulation is statistically equivalent, and provides a
statistical set of a little more than $10^9$ elements. We considered the
statistics that were obtained averaging over the full domain in one instant. The field
has been slightly modified in order to filter  out instantaneous
effects of the forcing, in other words,  a  turbulent kinetic  energy
inhomogeneity of about $20\%$ (in the spatial coordinate system). As
this bias was generated by the energy supply at the large scale
range, the two largest scales have been filtered out.
The resolved part of the energy spectrum extends up to $k\sim 330$.
The inertial range 
extends from $k\sim 10$ to $k\sim  70$, see  the compensated version of the 3D spectrum in figure \ref{spe3_compensated}. The higher wave-numbers, which are affected by the aliasing error, are not shown.

\begin{figure}
\begin{center}
\psfig{figure=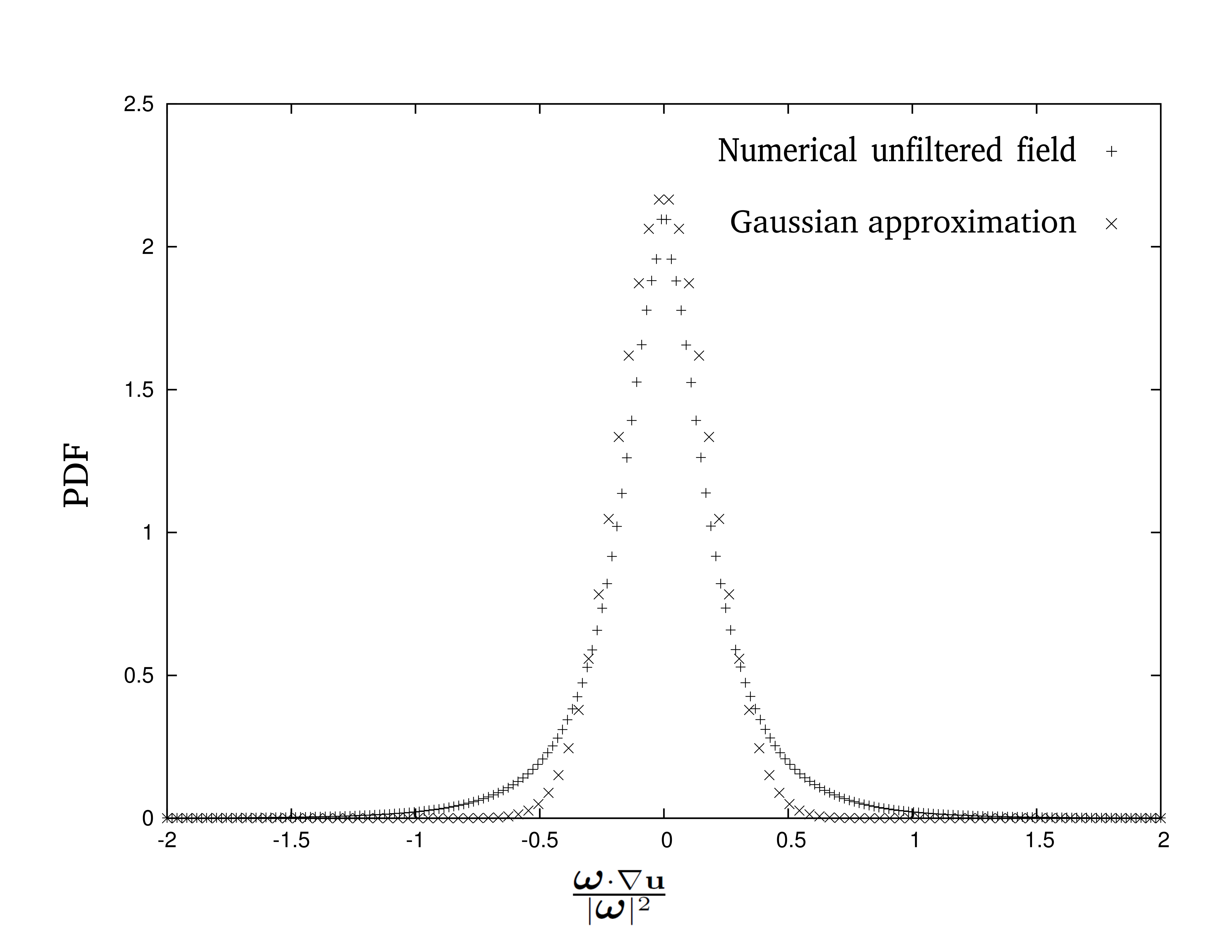,width=1\columnwidth}
\end{center}
\hskip -10mm
\vskip -10mm
\caption{Probability density function of one component of the vector $\frac{ \vor\cdot\nabla u }{|\vor|^2}$ in an isotropic velocity field with $Re_\lambda = 280$. Comparison with the Gaussian model. The Skewness is negligible, about $10^{-2}$. The Kurtosis however is very high and reaches a value of about $55$.}
\label{gaussiana}
\end{figure}

We focus now on  a few statistical
properties of $\frac{ \vor\cdot\nabla\UU }{|\vor|^2}$.
The pdf of the components of this vector (which are statistically equivalent,
since the field is isotropic) is shown in figure \ref{gaussiana}.
Symmetry with  the vertical axis is expected because of isotropy;
the skewness is in fact approximately $10^{-2}$, which is not meaningfully
far from zero. However,  the distribution cannot be approximated
with a \emph{Gaussian} function. In fact, the actual kurtosis is
approximately 55, which is very far from the Gaussian
value of 3.

The range of values attained by $f(\xx)$ is wide.
Values as high as a few hundreds were observed at a sparse spatial
points. In order to read the typical values of $f(\xx)$, we study
its \emph{survival function}. By denoting $F(s)=P(f(\xx)\leq s)$ the
cumulative distribution function  (cdf) of $f(\xx)$, 
the survival function is defined as the complement to $1$ of the
cdf,
\begin{equation}
S(s)=P(f(x)>s)=1-F(s).
\end{equation}

For each threshold $s$, $S(s)$
describes the \emph{probability that $f(\xx)$ takes  greater values
than $s$}.

\begin{figure}
\begin{center}
\psfig{figure=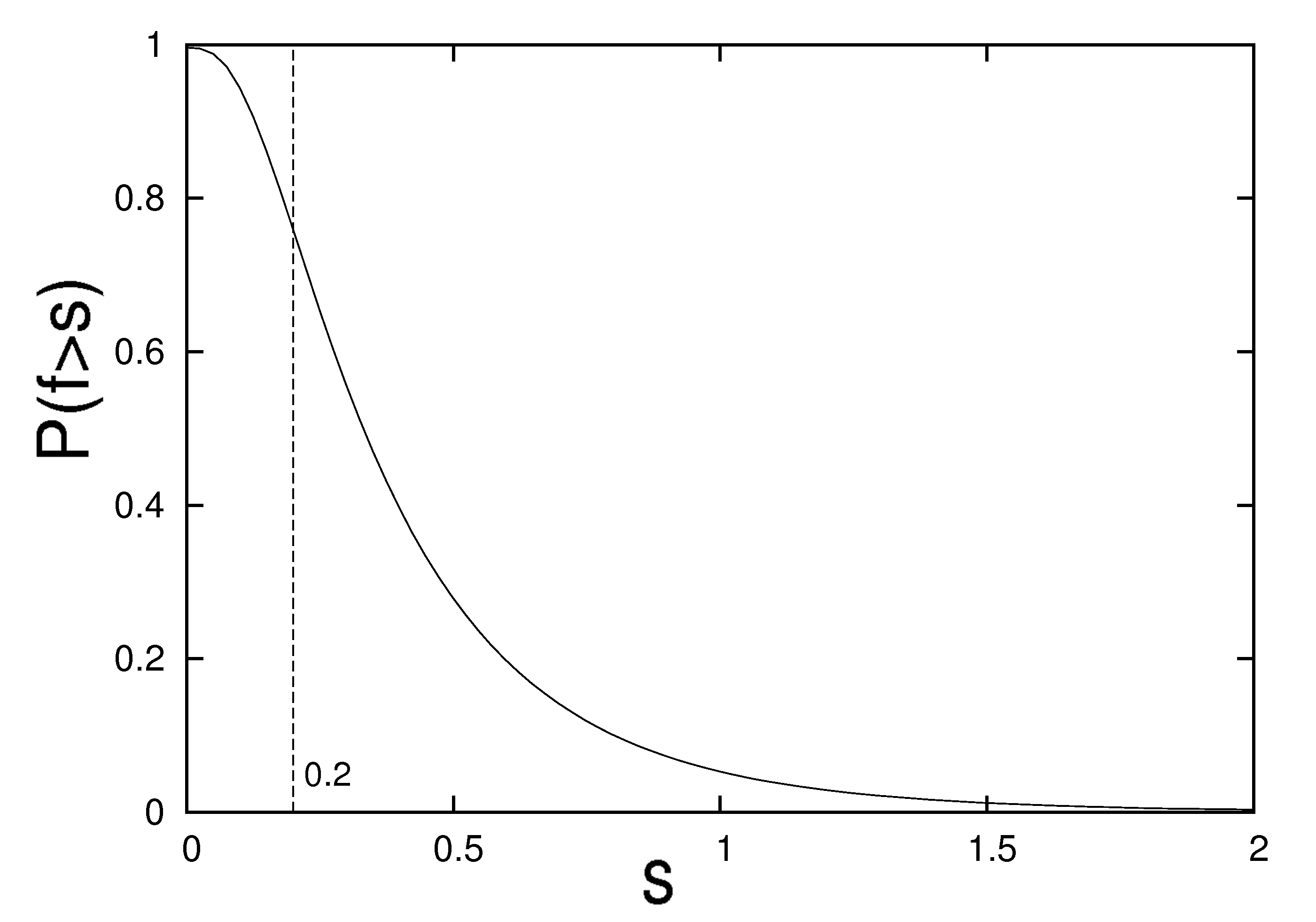,width=1\columnwidth}
\caption{Survival probability of the normalized stretching-tilting
function in a fully resolved isotropic  3D turbulent field
($P(f(\xx)\geq s)$, $Re_{\lambda}=280$. Unfiltered velocity field. The dashed vertical line indicates the value of $f$  where the probability density function is maximum.
} \label{confr_cdf}
\end{center}
\end{figure}

%
It has been found that, when $f(\xx)$ is evaluated on a well resolved isotropic turbulent field, the probability that $f(\xx)> 2$ is almost zero,
see figure \ref{confr_cdf}. Thus, $f(\xx)=2$ can be considered the maximum
statistical value that $f(\xx)$ can reach when the
turbulence is fully developed. 

\section{\label{Filtered}Properties of the survival function of the normalized stretching-tilting term: analysis on the anisotropically filtered field}

By means of suitable convolutions, the application of filters to the velocity field allows the behavior of the
function $f(x)$ to be studied in relation to the different turbulence scale ranges. This analysis is carried out using two spectral filters, a high pass and a band-stop filter. To  focus in an empirical way  on the  three principal kind of geometrical structures observed in turbulence, filaments, sheets and blobs, we use here a highly anisotropic kind of filter, which is less traditional than the axisymmetric-type filter. Of course, given the inadequacy of the spectral representation to account for the complex three-dimensional geometry of the turbulent structures commonly seen through visualization tools, the approach we use here is not rigorous and should be considered no more than propaedeutical.

\begin{figure}
\psfig{figure=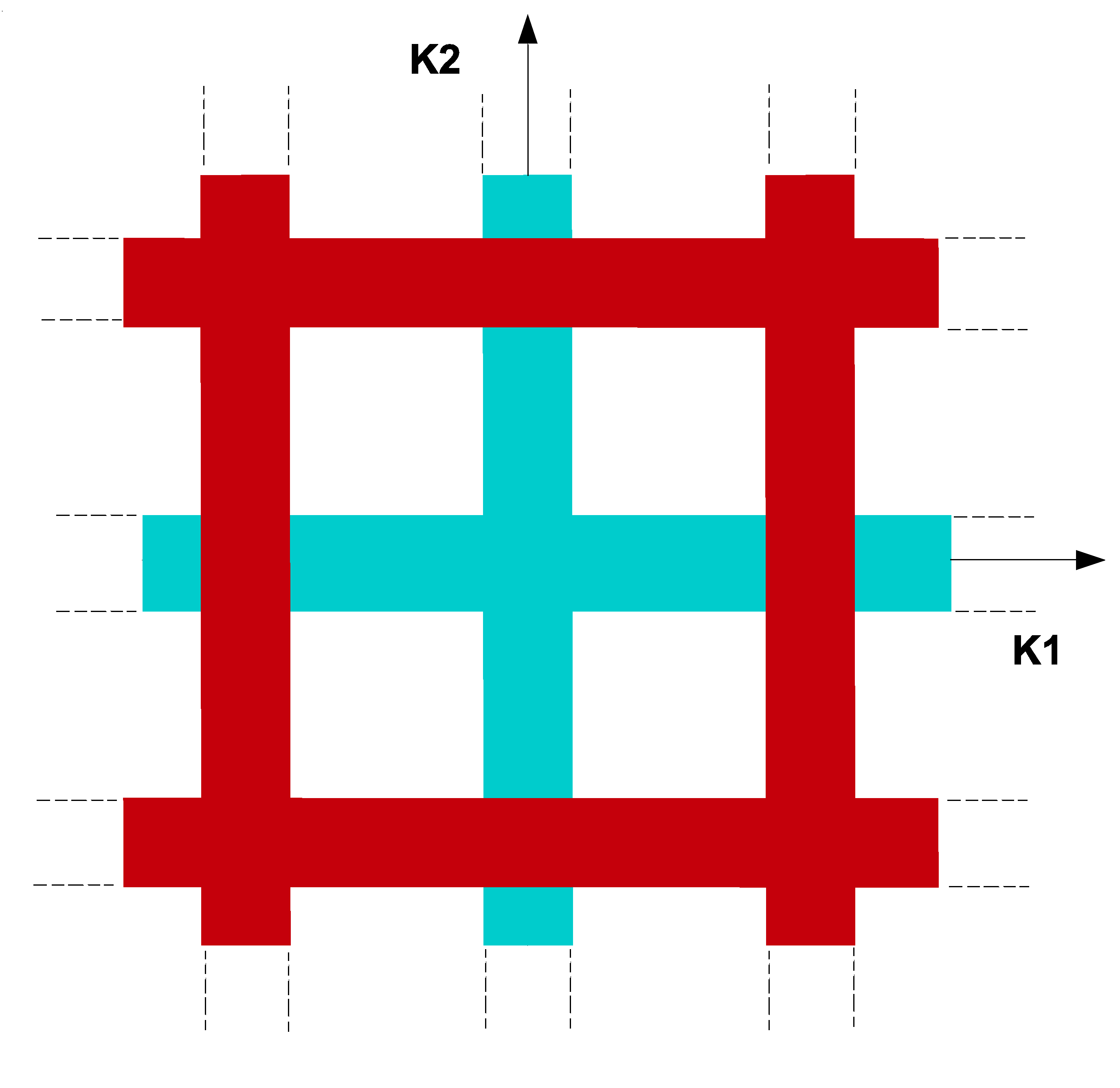,width=1\columnwidth}
\caption{Schema of the anisotropic filter here named as CROSS filter. Blue region: high-pass filter,
the wave-numbers under a certain threshold are partially removed, see eq. (4),
Red region:  band-stop filter, the wave-numbers inside a range are cut, see eqs. (5-7).}
\label{disegno_croce}
\end{figure}

The first filter is a sort of high-pass filter, which we refer to as
\emph{cross filter} and which allows  the contribution of
the structures that are characterized by at least one  \emph{large
dimension} to be removed. From the Fourier point of view, this means that the
structures whose wave-vector has at least one small component are
filtered out. One can here think about elongated structures as filaments and sheets or very large globular structures.  In figure \ref{disegno_croce}, a graphical
scheme of the filtering in the  wave number plane $k_1, k_2$ can be seen. The first filter we consider is here represented in blue and is a kind of high-pass filter which affects
all  wave-numbers that, along any  possible direction, have at
least one component under a certain threshold. Given the  threshold
$k_{MIN}$, the filter reduces the contribution of the modes with wave number components

\begin{equation}
k_1<k_{MIN}\,\, \textrm{ or }\,\, k_2<k_{MIN}\,\, \textrm{ or } k_3<k_{MIN}.
\nonumber
\end{equation}

The representation of this high-pass filter, $g_{hp}$,  can be given
by a function of the kind \cite{ti06}

\begin{equation}
\begin{split}
g_{hp}(\underline{k})=& \prod_i \phi(k_i; k_{MIN}), \; \phi(k_i, k_{MIN}) =\\
&{\displaystyle \frac{1}{1 + e^{-(k_i - k_{MIN})}}} 
\end{split}
\label{cf}
\end{equation}
\noindent Since function $g_{hp}$ filters any wavenumber that has at
least one component lower than the threshold $k_{MIN}$, it reduces
the kinetic energy of the filamentous (one component lower than
$k_{MIN}$), layered (two components lower than $k_{MIN}$) and blobby
(three components lower than $k_{MIN}$) structures. This filter is
efficient in reducing the integral scale of the turbulence
\cite{ti06}.



\begin{figure}
\begin{center}
\psfig{figure=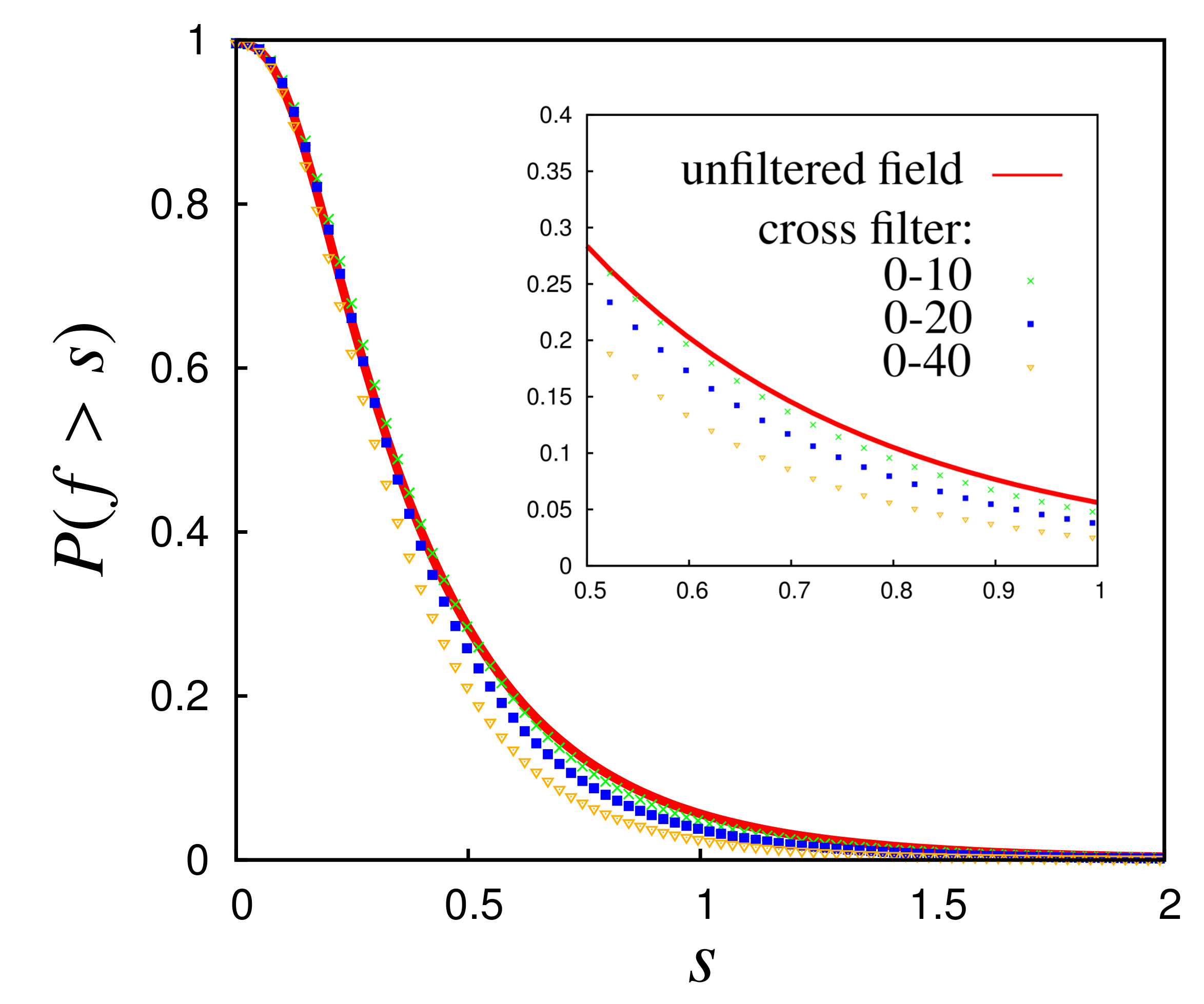,width=1\columnwidth}
\end{center}
\caption{
Survival probability of the normalized stretching-tilting function
in a high pass filtered isotropic  turbulent field. CROSS filter, see in fig.4 the blue region.}
\label{croce_passaalto}
\end{figure}

By varying the value of the threshold, $k_{MIN}$, it is possible to
consider different  scale ranges. The ranges  $0-10, 0-20, 0-40$ are
compared in figure \ref{croce_passaalto}. The first filtering affects the energy-containing range, while the other two
also include a part of the inertial range, which is visible in figure
\ref{spe3_compensated}.

The plots in figure 5 have  coherent behavior.
The  survival function $S$ for
the $0-10$ filtering is slightly below the values of  the
distribution of the unfiltered turbulence. This trend is confirmed
by the other two filterings, and the reduction grows as the
threshold $k_{MIN}$ increases. The high-pass filter has the  effect
of \emph{decreasing} the statistical values taken by $f(x)$ in the
domain. The wider the filtered range, the higher   the effect on
$f$.

It is possible to  say that when we reduce the weight of  the large-scale
structures (layers, filaments  or blobs), the local
stretching-tilting intensity decreases
with respect to the vorticity magnitude. On average, the values of
$f(x)$ go down. The wider  the range  affected, the lower
the probability value becomes. This  suggests that the large scales
contribute more to the stretching-tilting (the numerator of $f$)
than to the magnitude of the vorticity fluctuation (the denominator  of $f$). It should be noted that
this trend is consistent with the results in \cite{hsd08}. This
consistency also includes results relevant to the behaviour of the stretching fluctuation  and of the vorticity
magnitude in boundary layer  turbulence, see figures 6
and 9 in \cite{ah01}.  For the wider range $0 - 40$, a decrease of $30\%$ in
the cumulative probability is observed for a stretching-tilting of
about one half of the local vorticity. The decrease goes up to $80\%$
when statistically the stretching-tilting has the same magnitude of the vorticity, that is $f$ is close to 1,
see figure \ref{croce_passaalto}.

Let us now consider the behavior of $f(x)$ when the inertial and
dissipative ranges are affected by the filtering, namely a band-stop
filtering. In this case, the band width can be extended to obtain a low pass filtering.

This filter can be  obtained by reducing the contribution of a
variable band (see figure \ref{disegno_croce}, part in red)

\begin{eqnarray}
k_{MIN}<k_1<k_{MAX} &\textrm{  or  }& k_{MIN}<k_2<k_{MAX} \textrm{  or } \nonumber \\
k_{MIN}<k_3<k_{MAX}.
\nonumber
\end{eqnarray}

This yields the filter function $g_{bs}$
\begin{eqnarray}
g_{bs}(\underline{k})&=& \prod_i \overline{\phi}(k_i; k_{MIN}, k_{MAX}), \\
\phi(k_i; k_0) &=& {\displaystyle \frac{1}{1 + e^{-(k_i - k_0)}}},\nonumber \\
\overline{\phi}(k_i; k_{MIN}, k_{MAX}) &=& [1-\phi(k_i;k_{MIN})]+\phi(k_i; k_{MAX}) \nonumber
\label{bsf}
\end{eqnarray}

\begin{figure}
\begin{center}
\psfig{figure=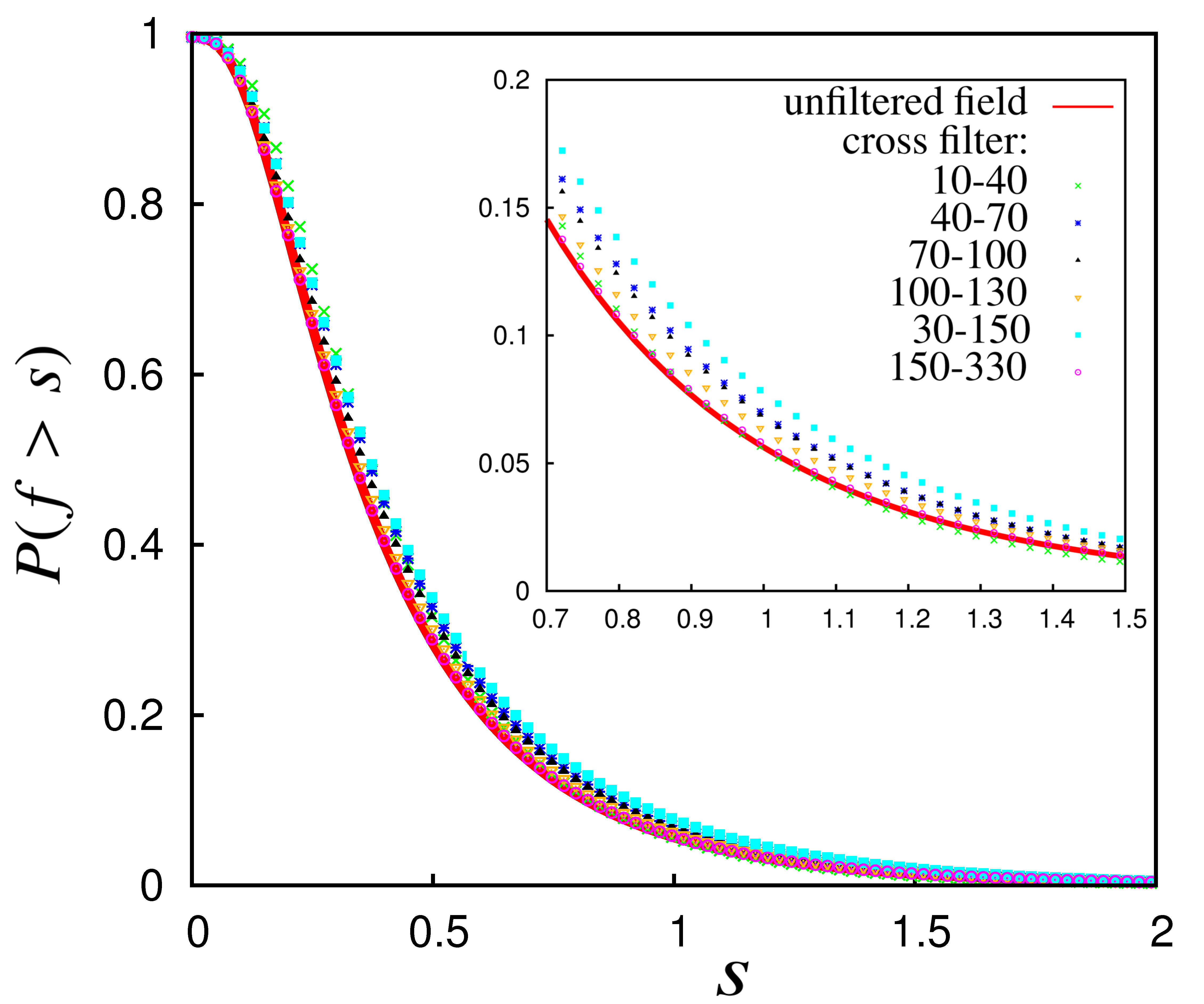,width=1\columnwidth}
\end{center}
\caption{Probability of the normalized stretching-tilting function
in a band pass  filtered isotropic turbulent
field of being higher than a threshold $s$. Control function: survival function
$1-F(x)$, band-stop filtering in various portion of the inertial range and in the dissipative range. }
\label{croce_range_inerz}
\end{figure}

The effects of the application of this band-stop filter on the
probability $P(f(\xx)\geq s)$ are  shown in figure \ref{croce_range_inerz}.

%

Let us  now consider the inertial range in  an \emph{extended way},
which includes the $-\frac{5}{3}$ range plus all the scales which
are not yet highly dissipative. 
The different bands are $10 - 40$, large scale inertial filtering,
$40 -70$ intermediate scale inertial filtering, $70 - 100$ small
scale
inertial filtering, $100 - 130$ near dissipative, 
$30 - 150$ intermediate-inertial/small scale  filtering, $150 - 330$
dissipative scale filtering.
 Once again all the filtered ranges  induce the same effects: for
$s < 1/2$,  a slight \emph{increase} in the survival
probability.
For  small values of $s$, the most effective filtering (i.e.
the ones which produce the highest increase) is  the
$10 - 40$, while, for higher statistically relevant
values, $ 0.5 < s < 2$, the most effective result is
obtained filtering over the whole inertial range, $ 30 < k < 150$.
In this case, an increase of about $60\%$ is observed for $s = 1$ and of about
$80\%$ for $s = 1.5$

This highlights the fact that the structures of the inertial
range contribute more to the intensity of the vorticity field than
to stretching and tilting. The general trend
is almost inverted with respect to the case of the high pass filtered
turbulence (compare the 0-40 and 10-40 results in figures 4 and 5,
respectively) and this can be confirmed, with slight differences, as long
as we enlarge the amplitude of the filtering band to get closer to
the  dissipative range. Finally, moving toward the dissipative range ($150 < k < 330$),
the band-stop filter becomes a sort of low-pass filter. By filtering these wave numbers, the  obtained effect is minimum,
although we have removed the contribution of more or less the
highest 200 wave-numbers (see figure \ref{croce_range_inerz}).

\begin{figure}
\includegraphics[width=0.28\textwidth]{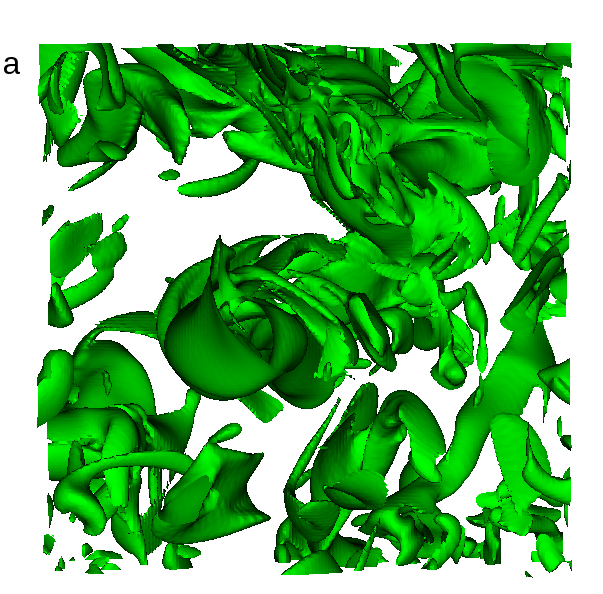}
\includegraphics[width=0.2835\textwidth]{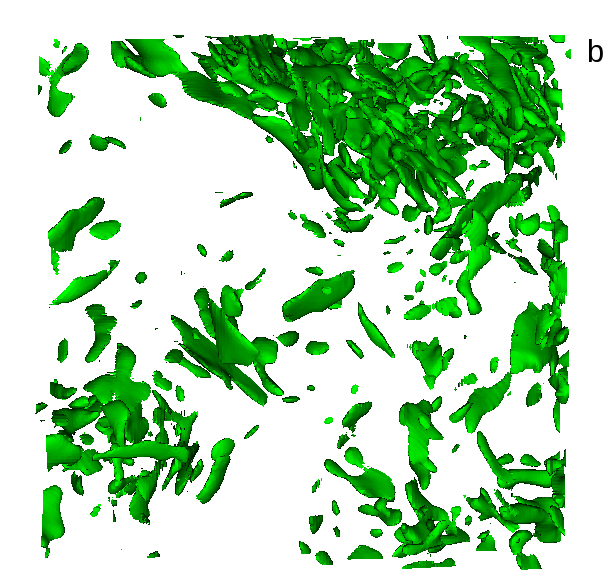}
\includegraphics[width=0.2975\textwidth]{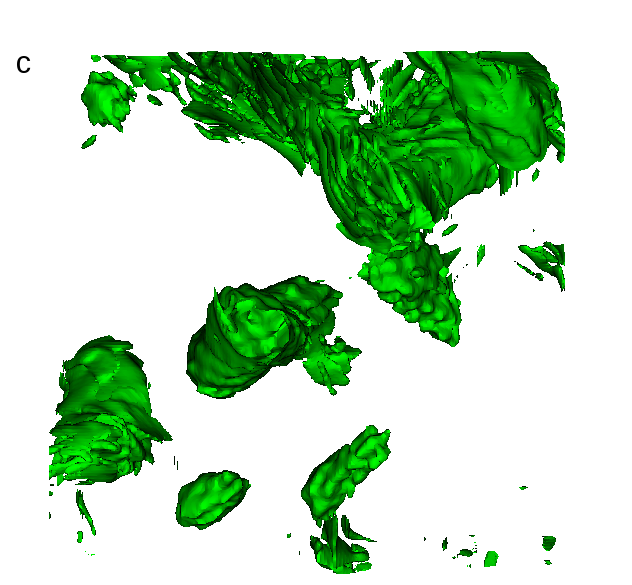}
\includegraphics[width=0.2695\textwidth]{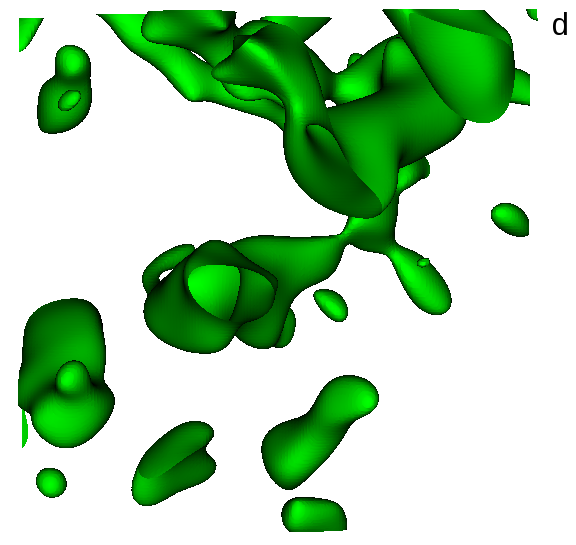}
\caption{\label{omega_1_comp} Three dimensional visualization of the surfaces where one  vorticity component has the value $17.5 sec^{-1}$. The root mean square value of the vorticity magnitude in the field is $|\omega|_{rms} = 30.3 sec^{-1}$. As reference, we consider a water field  with $Re_\lambda = 284$, $u_{rms} = 0.09 m/sec$, $\nu=0.96*10^{-6} m^2/sec$ (water viscosity at $23$ $^{o}C$), Taylor micro-scale $\lambda = 3 mm$ and integral scale $l = 56 mm$. Panel (a): unfiltered field; panel(b): the wave number range 0-20 is filtered out by using the high-pass cross filter, panel (c): the wave number range 30-150 is filtered out by using the band-stop cross filter;  panel (d): the wave number range 30-infinity is filtered out by using the low-pass cross filter, i.e. by letting $k_{MIN} \rightarrow \infty$, see figure \ref{disegno_croce}. The visualization shows a $256^3$ portion of the numerical field simulated on a $1024^3$ point grid. The field is visualized by means of VisIt (https://wci.llnl.gov/codes/visit/).}
\end{figure}

To see the effect of the filters on the vortical structures, the vorticity magnitude  has been visualized in two ways. The first is a  volume rendering of the surfaces where one of the vorticity components is close to the root mean square value, see figure \ref{omega_1_comp}. The instantaneous field we are considering is homogeneous and isotropic, so all the components are statistically alike and  it suffices to observe one  component only.
In  panel (a) the unfiltered field is visualized and a complex picture made of an elongated, thick, sleeve-like structures which are enfolded and twisted can be seen. In panel (b) the structures with at least one wavenumber component below 20 are smoothed out, see equation (\ref{cf}). One here sees a more sparse distribution of mostly elongated and nearly flat structures with a much shorter length with respect to panel (a). The mutual folding and twisting seems reduced. 
This image is related to the survival distribution in figure 5, where a depression of stretching-tilting over the vorticity magnitude is reported for ratio values above 0.3.
Panel (c) shows the band-stop filtered field, where wavenumbers in between 30 and 150 are smoothed out. Basically,  most of the inertial and larger dissipative structures are removed. Here, it is interesting to observe that the surface is much more corrugated that in panel (a), which leaves a definitive view of the small scale above  wavenumber 150.  Some of these structures are elongated, others are globular. The image does not discourage the idea that the large and the very small scales directly interact. Finally, in panel (d) one sees the structures that have at least one wavenumber component in the range 0 - 30. Here, large unruffled structures which are mainly globular can be seen. The images in panel (c) and (d) represent instances of the statistical  situation described in figure 6,  where the survival ratio of the stretching and vorticity intensities is enhanced with respect to the natural situation. Thus, it seems that the partial absence of the inertial range amplifies the stretching-tilting process.  It should also be noted that the structure spatial  density is distributed in almost the same way  in all the panels, though the density levels are different.

\begin{figure}

\includegraphics[width=0.5\textwidth]{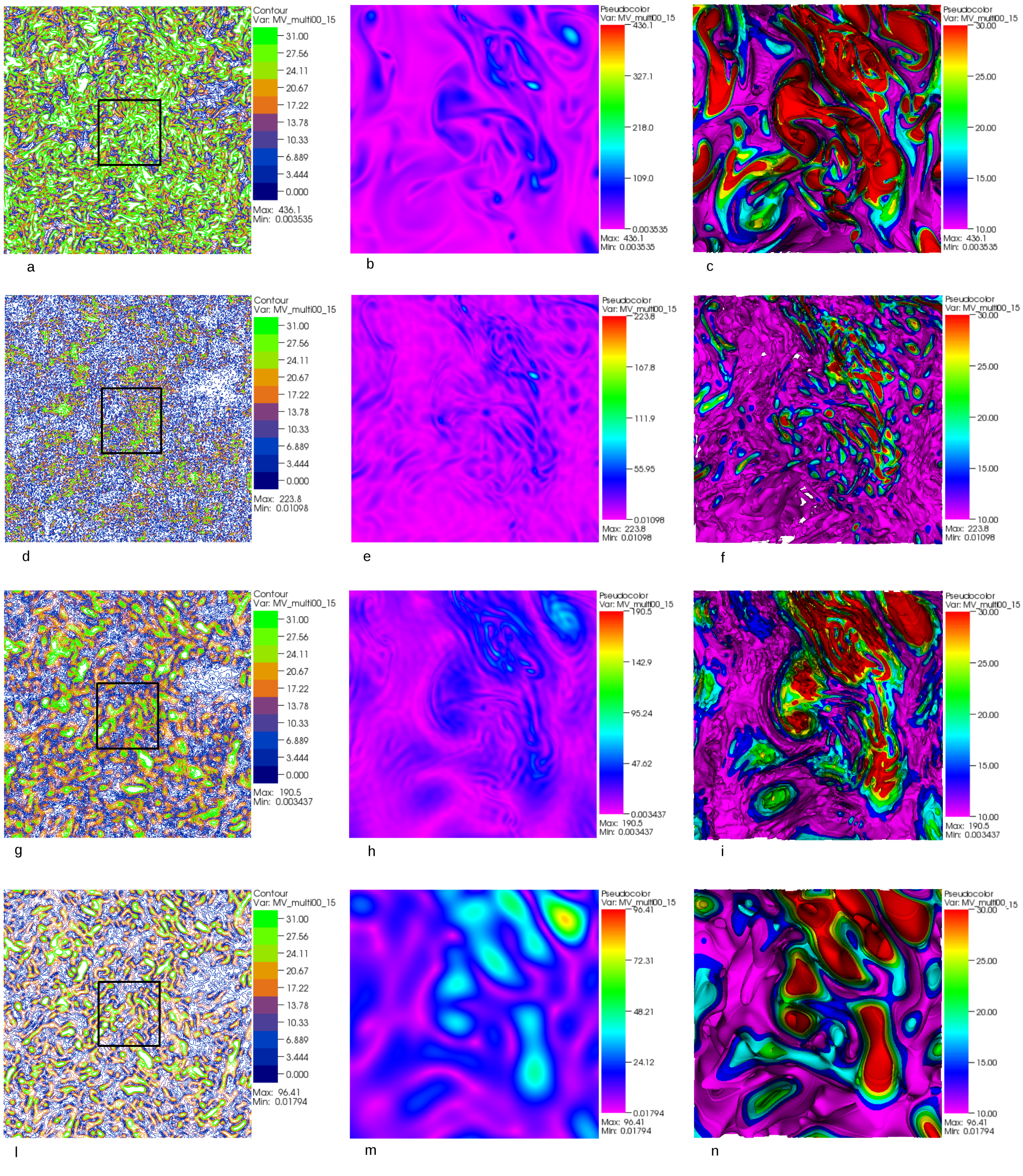}
\caption{\label{all} Visualization of the vorticity magnitude in a section parallel to one face of the computational box. Reference  isotropic turbulence:  $Re_\lambda = 280$, $u_{rms} = 0.09 m/sec$ (rms of the velocity), $\nu=0.96*10^{-6} m^2/sec$ (water viscosity at $23$ $^{o}C$), Taylor micro-scale $\lambda = 3mm$ and integral scale $l = 56 mm$, $|\omega|_{rms} = 30.3 sec^{-1}$ (root-mean-square of the vorticity magnitude), $1024^3$ grid domain points. First row (a,b,c): unfiltered field. Second row (e,f,g): the wave number range 0-20 is filtered out by using the high-pass cross filter. Third row (g,h,i): the wave number range 30-150 is filtered out by using the band-stop cross filter. Fourth row (l,m,n) the wave number range 30-infinity is filtered out by using the low-pass cross filter. It is possible to see: in the first column (a,d,g,l) the vorticity magnitude countourplots  of the entire $1024^3$ grid domain. In the second and third column (b,e,h,m and c,f,i,n) Pseudocolor plots of a $256^3$ portion of the grid (black box in the previous column) are shown. In the third column the range of magnitude values  in between 10 and 30 is visualized to show the details of the part of the field where the vorticity is below its rms value. The part above the rms is in fact visualized in the central column, where all the range of values is included. The Pseudocolor method maps the data values of a scalar variable to color. The plot then draws the colors onto the computational mesh. The field is visualized by means of VisIt (https://wci.llnl.gov/codes/visit/)}.
\end{figure}

We have tried to visualize the  filtering effect also by means of contour plots and pseudo color imaging of the vorticity magnitude in a flat section of the $1024^3 field$, see figure \ref{all}. Here, the lines in the images in the left column are the contours of iso-surfaces of the  vorticity magnitude. Starting from the top, one sees the unfiltered field, the high pass filtered field (the wavenumbers above 20 are kept), the $30-150$  band-stop, and the low pass filtered field (the range $30- infinity$ is removed).  The contour plot technique is very popular, but, apart from clearly  showing the reduction of the turbulence scale size when the large scales are missing, it does not give much information. Essentially, structure contours  appear to have the peanut shape which typically hosts vortex dipoles.

In the enlarged views in the central and right columns, the pseudocolor plots are richer in information. Where the larger scales are removed, panels (e) and (f), the survival probability of intense  stretching is reduced. The range of variation for the vorticity magnitude is very large, the root mean square value is about  seven times smaller than the maximum value. When only the large scales are left, panels (m) and (n), the rms value is about one third of the maximum value and the stretching is enhanced.  When the inertial scales are removed, the large scale appears to be wrapped by the small scales. The big dipoles are surrounded by thin wavy-like sheets and the small scales are attached to the large ones. In this situation it is not possible to neglect their direct interaction. In regions where a large scale is missing, the small scales are also missing. Viceversa, a low level of direct interaction between the largest and the smallest scales would have been confirmed, if small scales  had been sparsely distributed in regions where the large scales are not present and  would not have surrounded the large structures in regions where the latter are present.

To complement the understanding of the visualization in relation to the anisotropic filtering here used, we have considered the alignment between the eigenvectors of the strain rate tensor and the direction of the vorticity.
Function $f$ in fact, see equation (1), can also be written as $f(\xx)=\frac{|S|}{| \vor|}[s_i^2 (e_i * e_\omega)^2 ]^{1/2}$ where $|S|$ is the magnitude of the strain rate tensor,  $s_i$ are the eigenvalues of the strain rate tensor $S_{ij}$ normalized by $|S|$, $s_i = \lambda_i/|S|$,  and  $(e_i*e_\omega)$ describe the alignments between the eigenvectors of $S_{ij}$, denoted $e_i$, and the direction of  the vorticity $e_\omega$. Figure \ref{grafico_cosdir}  shows the probability density function of these alignments in the reference unfiltered field and in two filtered cases described in this work: the  high pass filter where the smallest $20$ wavenumbers are removed, see equation (3) and the band-stop filter where wavenumbers  $30-150$ are removed, see equation (4). We first observe that the standard trend of alignments is not fully spoiled by  the filtering. In both filtering cases, eigenvector 2 reduces its alignment, while eigenvector 3 reduces its misalignment. Conversely, eigenvector 1 instead shows a different behavior. In the band-stop filtering case (large scale dominate) eigenvector 1  slightly increases the alignment. In the high-pass filtering, eigenvector 1 reduces the alignment that becomes statistically equal to that of the eigenvector 3.  This is confirmed, see Table 1, by considering the ratio among the field averaged strain rate tensor eigenvalues and related component of the  enstrophy production,  $<\sigma_i>/<\sigma_{tot}>$, where $< \cdot >$ is the average over all the computational domain ($1024^3$ grid point box), $\sigma_i=\omega^2\lambda_i cos^2(e_\omega,e_i)$ and $\sigma_{tot}=<\sum_i\omega^2 \lambda_i cos^2(e_\omega,e_i)>$. All filters increase the gap between the eigenvalue $<\lambda_1>$ and $<\lambda_2>$ and the gap between $<\lambda_2>$ and $<\lambda_3>$. However, when the large scales are missing, eigenvalues 1 and 3 are very close in modulus. The same happens to the modulus of their related  enstrophy production components. When the inertial scales and part, or the entire, dissipative range are removed the mutual relation among the eigenvalues and the modulus of the production components changes less with respect to the natural turbulence.

In all the three cases the filtering reduces the average values of the largest and intermediate eigenvalue, $<\lambda_1>$  and $<\lambda_2>$.


\begin{figure}
\includegraphics[width=0.5\textwidth]{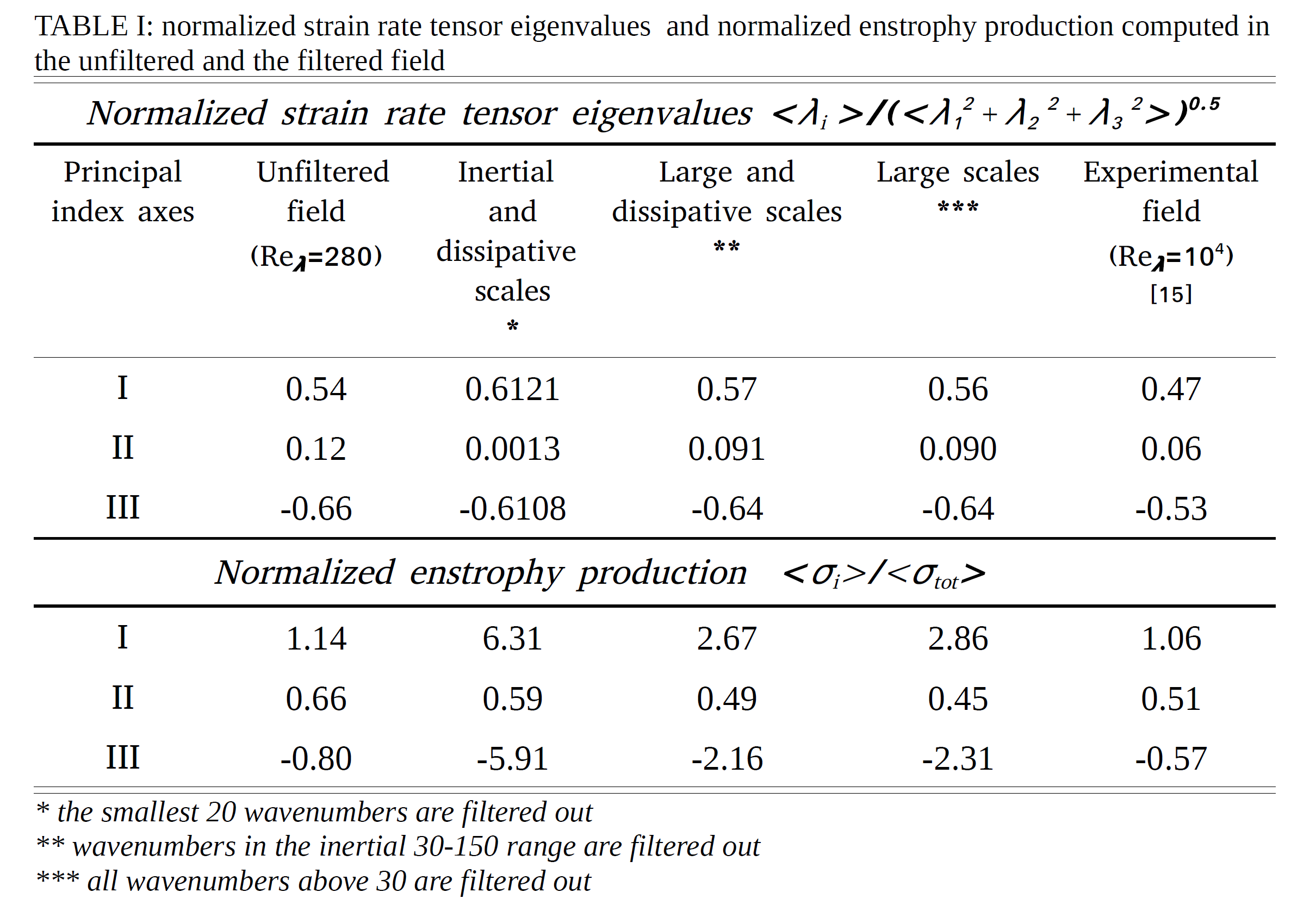}
\end{figure}

\begin{figure}
\includegraphics[width=0.5\textwidth]{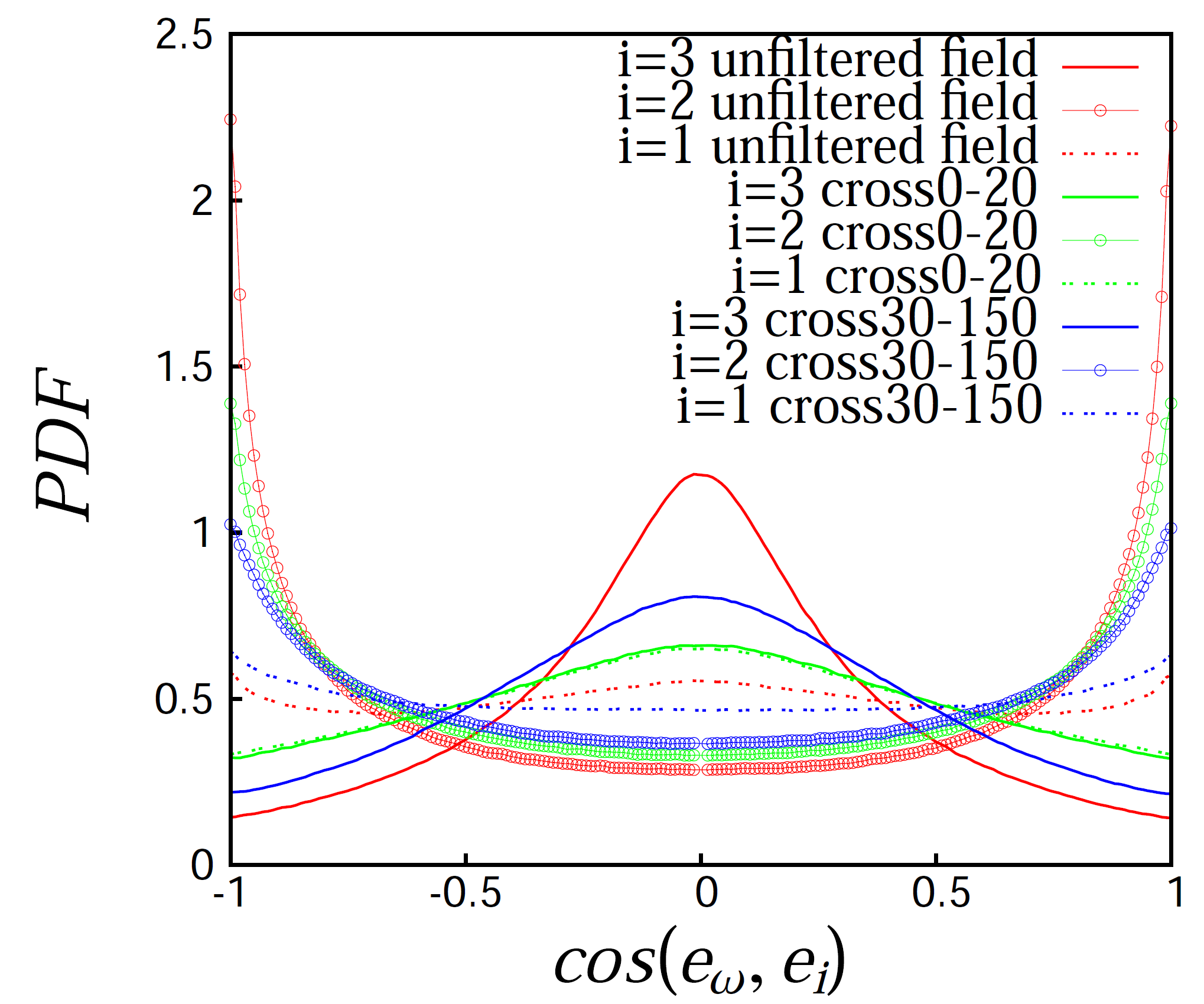}
\hskip -5mm
\vskip -5mm
\caption{\label{grafico_cosdir}PDFs of the cosine of the angle between vorticity (\vor), and the eigenvector, $e_i$, of the rate of strain tensor ($S_{ij}$). The red lines refer to unfiltered field; the green lines refer to the wave number range 0-20 filtered out by using the high-pass cross filter; the blue lines refer to the wave number range 30-150 filtered out by using the band-stop cross filter.}
\end{figure}


At this point, let us consider the dual nature of the filaments and
sheets,  as regards their inclusion  in the categories of the small  and large scales.  A filament which is filtered out by
the filter $g_{hp}$  because  it has a small wave number (the axial
wave number component), will  have two large wave number components
(the ones normal to the filament axis). Due to these wave components, it will also be filtered
out  by the filter $g_{bs}$. A similar situation also holds for
the sheets. Thus  filaments and sheets are always partially removed when a filtering, either anisotropic or isotropic, is applied.
The situation changes with compact structures (the blob), which  non ambiguously belong either to the large scale
range or to the intermediate-small scale range. The different
behavior shown in figures \ref{croce_passaalto} and \ref{croce_range_inerz} is therefore mainly due to the blob
contributions, and, since the variation in the cumulative
distributions is opposite and almost of the same magnitude, it is possible to deduce that the partial removal of the background filaments and
sheets, which is always done regardless of the filter typology (high pass, band
pass, low pass,..), is not statistically relevant. In other
words, it appears reasonable to conclude that until a better way to select and remove anisotropic structures such as filaments and sheets is found, first order statistical modification associated to their presence/absence will not be clearly seen.

Lastly, it is interesting to observe that box filtering  small
scales modifies the stretching statistics to a great extent. A field
filtered in such a way shows a finite probability of having more than twice the amount of stretching/tilting compared to the enstrophy \cite{tim07, tim08}. In the context
of the Large Eddy Simulation methodology, where this kind of filtering is commonly used,  it is possible to deduce that, when a
fluctuating field shows such a feature,  the field is unresolved.
As a consequence, it is possible to build a criterion that locates
the regions of the field where the inclusion of a subgrid term in
the governing  equations is advisable (see also \cite{timm13}).

\section{\label{Shell} Conclusions}

To summerize, we have collected a set of statistical information
about the stretching and tilting intensity of vortical structures normalized by the enstrophy,
$f(\xx)=\frac{| \vor\cdot\nabla\UU |}{|\vor|^2}(\xx)$,
in  isotropic turbulence. 
A first result is that there is  a very small probability of having a larger stretching/tilting  of intensity than the double 
of the square of the vorticity magnitude. Then, if  compact structures (blobs) in the
inertial
range  are filtered out, it can be seen that the probability of having higher $f$
 than a given threshold $s$  increases by 20\% at
$s=0.5$, and  by 60-70\% at $s=1.0$. If, on the other hand, larger blobs  are
filtered,  an opposite situation occurs. 
The unfiltered field
is  thus a separatrix  for the  cumulative probability function. This behavior - high
fluctuation vorticity magnitude $\rightarrow$ low stretching,
and viceversa - agrees with general aspects highlighted by a number of laboratory and numerical analyses, \cite{ag05,hsd08} also in near wall turbulent flow configurations\cite{Tsinober_92_jfm}.  The present
observations need to be associated to the non discriminating effect of
filtering  on  filaments and sheets, which is due to their  specific nature that cannot be reconciled inside
either a category of  small or large scales.  It has also been shown that a high intermittency is associated to $f$, whose kurtosis  is as high as 55.

The probability density function of the alignments between the strain rate eigenvectors and the vorticity is in part modified by the anisotropic filtering here investigated. In particular, we observe that, though the standard trend of alignments is not fully spoiled,  eigenvector 2 reduces its alignment, while eigenvector 3 reduces its misalignment. Conversely eigenvector 1 shows a different behaviour. In the band stop filtering case (large scale dominate) eigenvector 1 slightly increases the alignment. In the high pass filtering case (inertial scales dominate), eigenvector 1 reduces the alignment that becomes statistically equal to that of the eigenvector 3. This is confirmed by considering the mutual ratio among the averaged strain rate eigenvalues and related components of the  enstrophy production. Both filters increase the gap between the most extensional eigenvalue $<\lambda_1>$ and the intermediate one $<\lambda_2>$ and the gap between this last $<\lambda_2>$ and the contractile eigenvalue $ <\lambda_3>$. However, when the large scales are missing, the modulus of eigenvalues 1 and 3 become nearly equal, similar to the modulus of the related components of the enstrophy production.



\section{\label{} Acknowledgments}

We are grateful to Professor James J. Riley for  fruitful discussions.  We thank the referees for their constructive
reviews  and the many suggestions.  This work
was carried out in cooperation with the International Collaboration for Turbulence
Research.


\end{document}